\documentstyle[twocolumn,aps,epsf]{revtex}

\begin{document}
\draft
 \twocolumn[\hsize\textwidth\columnwidth\hsize\csname @twocolumnfalse\endcsname

\title{ Effects of substituting rare-earth ion $R$ by non-magnetic 
impurities in $R_2BaNiO_5$ - theory and numerical DMRG results
}
\author{T.-K. Ng}
\address{
Department of Physics,
Hong Kong University of Science and Technology,\\
Clear Water Bay Road,
Kowloon, Hong Kong
}
\author{Jizhong Lou, Zhao-Bin Su}
\address{
Institute of Theoretical Physics,
P. O. Box 2735,
Beijing 100080, P.R. China
}
\date{ \today }
\maketitle
\begin{abstract}
   In this paper we study the effect of substituting $R$ (rare-earth ion)
by non-magnetic ions in the spin-1 chain material $R_2BaNiO_5$. Using a
strong-coupling expansion and numerical density matrix renormalization group calculations, we show that spin-wave bound states are formed at the impurity
site. Experimental consequences of the bound states are pointed out.

\end{abstract}

\pacs{PACS Numbers: 75.10.-b, 75.10.Jm, 75.40.Mg}

]

\narrowtext

   The antiferromagnetic Heisenberg spin chains has been a subject of 
immense interest since Haldane pointed out that the low energy physics of
integer and half integer spin chains are fundamentally different\cite{hal}. 
With advance in both experiments\cite{e1,e2} and numerical 
techniques\cite{n1}, the prediction by Haldane is now generally accepted. 
More recently, a series of experiments on the family of 
quasi-one-dimensional materials with a general formula
$R_2BaNiO_5$\cite{e2,e3,e4,e5,e6,e7,e8,e9} where $R$ is one 
of magnetic rare-earth elements substituting fully or partially $Y$ have 
made it possible to study the effect of a staggered magnetic field on a 
$S=1$ spin chain. The compound $Y_2BaNiO_5$ is believed to be a good 
realization of $S=1$ Haldane gap system\cite{e2}. The $S=1$ spins are 
coming from the $Ni^{2+}$ ions. Other members have additional magnetic 
ions $R^{3+}(R\neq{Y})$ positioned between two neighboring $Ni$ chains. 
The coupling between these ions and $Ni^{2+}$ ions are weak. Nevertheless, 
these magnetic ions are AF ordered below certain N\'eel temperature $T_N$, 
and the ordering affects the $Ni$ chains by providing them an effective
background staggered magnetic field. It is expected that the three modes 
of the Haldane triplet will be splitted by the staggered field at $T<T_N$, 
with the longitudinal mode energy increases roughly twice as fast as the
transverse ones with decreasing temperature below $T_N$\cite{t1,t2,n2,t3}. 
These effects were indeed observed in neutron scattering
experiments\cite{e7,e8,e9}. 

   In this paper we shall study the effect of replacing $R$ ions by 
non-magnetic impurities. In the case of pure $S=1$ Haldane spin chains,
the effect of replacing a spin by non-magnetic ion has received much 
attention because of the theoretical prediction of $S=1/2$ excitations 
localized at the ends of broken $S=1$ spin chains\cite{t2}. The existence 
of end excitations with fractional spin magnitudes are believed to be a 
general property of Heisenberg spin chains and is a consequence of the
topological properties of these systems\cite{ng}. However, the effect 
of substituting $R$ by non-magnetic impurities is very different. As
far as the $Ni^{2+}$ chain is concerned, the replacement of an $R^{3+}$ 
ion by an non-magnetic impurity just alters the local effective 
(staggered) magnetic field seen by the spin chain, and the topological 
character of the spin chain is unaffected. Therefore, topological 
$S=1/2$ end excitations are not expected to be induced by the $R$-ion
substitutions. Nevertheless, we shall show in the following that 
although topological end excitations are not induced by $R$-ion 
substitutions, $S=1$ magnon bound states are formed around the
impurity and give rise to observable effects on the system. 

   Our starting point is the effective Hamiltonian for the $Ni$ chain of 
the $R_2BaNiO_5$ family at $T<T_N$,
\begin{equation}
\label{ham}
H_{bulk}=J\sum_i\vec{S}_i.\vec{S}_{i+1}+h\sum_i(-1)^iS_i^z,
\end{equation}
where $J$ is the exchange constant, $h=gS\mu_BH_s$ where 
$H_s\sim{M}_s$ is the amplitude of the staggered field and $M_s\neq0$ 
at $T<T_N$ is the sublattice magnetization from the $R$ ions. Replacing 
an $R$ ion by non-magnetic impurity at site $i=0$ generates an effective 
local magnetic field $h'$. We shall assume a corresponding impurity 
Hamiltonian 
\[ 
H_{im}=h'S_0^z,   \]
in the following, where $h'=-h$ for non-magnetic impurity. We shall 
consider $h'$ as an arbitrary parameter, and study it's effect on the 
spin chain in the following. The Hamiltonian $H=H_{bulk}+H_{im}$ will be 
studied using two different approaches. In the first approach, we 
replace $H$ by the corresponding lattice non-linear-$\sigma$-model (or
coupled-rotor model)\cite{chn}, and shall study the problem in the
strong-coupling limit when the Haldane gap $m_H$ is much larger than 
the coupling between rotors. This approach provides a rather simple 
physical picture on the effects of staggered magnetic field and 
impurity on the Haldane chain. The results obtained from the 
strong-coupling theory is compared with results obtained directly 
from diagonalizing $H$ with the numerical density matrix 
renormalization group (DMRG) method. Combining the results from the 
two methods we believe that our study provides a clear picture on 
the effect of substituting $R$ by non-magnetic impurities in the 
$R_2BaNiO_5$ compound. First we consider the coupled-rotor model.

   In this treatment the low energy physics of an integer Heisenberg 
spin chain in staggered magnetic field $h$ is described by 
the Hamiltonian\cite{chn}
\begin{equation}
\label{hrotor}
H_{rot}=\Delta\sum_i\vec{L}_i.\vec{L}_i-J'\sum_i\vec{n}_i.\vec{n}_{i+1}
-h\sum_in^z_i,
\end{equation}
where $\vec{n}_i$ is a unit vector located at the link between site 
$i-1$ and site $i$, and $\vec{L}_i$ is the angular momentum 
operator conjugate to $\vec{n}_i$. $\Delta$ and $J'$ are parameters 
to be determined. In the strong-coupling expansion, we assume 
$\Delta>>J'$ and treat $J'$ as a small perturbation. In particular, 
to study the one-magnon excitations, we shall only keep the $l=0,1$ 
states ($l$'s are eigenstates of angular momentum operator 
$\vec{L}.\vec{L}$) and shall diagonalize the Hamiltonian $H_{rot}$ 
approximately in this subspace. We shall neglect renormalization of 
ground state by two magnon processes in our theory. First 
we consider the limit when $h=0$.

  After some straightforward algebra, we obtain the effective 
Hamiltonian in the $l=0,1$ subspace, 
\begin{equation}
\label{Heff0}
H_{eff}^{(0)}=\Delta\sum_{i,m}l_i(l_i+1)-{J'\over3}\sum_{i,m}(a_{im}^+
a_{i+1m}+a_{i+1m}^+a_{im}),
\end{equation}
where we have set $\hbar=1$. $l(l+1)=0,2$ are the eigenvalues of 
$\vec{L}.\vec{L}$ in the $l=0,1$ subspace. $m$ is the eigenvalue of
$l_z$, $a^+_m(a_m)$ are raising and and lowering operators with 
matrix elements $<1m|a^+_m|00>=<00|a_m|1m>=1$, and with all other 
matrix elements equal to zero. $|lm>$ are the angular momentum 
eigenstates. The one-magnon eigenstates of \ (\ref{Heff0}) are 
triplets $(l=1, m=0,\pm1)$ with  energy spectrum
$\epsilon_k=2(\Delta-(J'/3)cos(k))$, and with Haldane gap 
$m_H=2(\Delta-J'/3)$.

  We now consider the staggered magnetic field $h$. First we consider 
the effects of an effective staggered magnetic field $h_{eff}$ on a 
single rotor. The term $h_{eff}n^z$ couples the 
$|00>$ and $|10>$ states and the new eigenstates of a single rotor 
in the $l=0,1$ subspace are  
\begin{eqnarray}
\label{eigr}
|G> & = & \beta|1,0> + \alpha|0,0>,\;   \;\;\; 
|0> = -\alpha|1,0> + \beta|0,0>,     \nonumber \\
|1> & = & |1,1>,\;    \;\;\;
|-1> = |1,-1>,
\end{eqnarray}
where $|G>$ is the new ground state, and $|m> (m=0,\pm1)$ are the new
excited states with $l_z=m$. The coefficients $\alpha$ and $\beta$
are given by
\begin{equation}
\label{coeff}
\alpha^2={1\over2}(1+\gamma(h_{eff})^{-1})\;, \;\;\;
\beta^2={1\over2}(1-\gamma(h_{eff})^{-1}),
\end{equation}
 where $\gamma(h_{eff})=\sqrt(1+{h_{eff}^2\over3\Delta^2})$ and with
corresponding eigen-energies $E_G=\Delta(1-\gamma(h_{eff}))$,
$E_0-E_G=2\Delta\gamma(h_{eff})$, and $E_{\pm1}-E_G=\Delta
(1+\gamma(h_{eff}))$. Notice that the energies of the longitudinal 
($m=0$) and transverse ($m=\pm1$) magnon modes are splitted by 
$h_{eff}$. The staggered magnetic field also induced a non-zero 
ground state expectation value of staggered magnetization,
$<n_z>=<G|n_z|G>={2\over\sqrt{3}}\alpha\beta$. To proceed further
we replace the rotor Hamiltonian \ (\ref{hrotor}) by a mean-field
rotor Hamiltonian,
\begin{equation}
\label{hmf}
H_{rot}^{(MF)}=\Delta\sum_i\vec{L}_i.\vec{L}_i-J'\sum_i\vec{n}'_i.
\vec{n}'_{i+1}-h_{eff}\sum_in_i^z,
\end{equation}
where $\vec{n}_i=<n_z>+\vec{n}'_i$ and $h_{eff}=h+2J'<n_z>$. Treating 
the $J'\vec{n}'_i.\vec{n}'_{i+1}$ term as perturbation in the $l=0,1$ 
subspace as before, we obtain after some straightforward algebra 
the mean-field equation
\begin{equation}
\label{mfe}
1-{h\over{h}_{eff}}={4J'\over3\Delta}\gamma(h_{eff})^{-1},
\end{equation}
which determines $h_{eff}$ for given $h$, and with corresponding 
eigenvalue spectrum for the one magnon states,
\begin{eqnarray}
\label{spectrum}
\epsilon_{\pm1}(k) & = & \Delta(1+\gamma(h_{eff}))-{2J'\over3}
\alpha^2cos(k),   \\   \nonumber
\epsilon_{0}(k) & = & 2\Delta\gamma(h_{eff})-{2J'\over3}
(\alpha^4+\beta^4)cos(k),
\end{eqnarray}
for the transverse ($m=\pm1$) and longitudinal ($m=0$) magnons,
respectively. At small $h$, it is easy to show from \ (\ref{mfe}) that
$h_{eff}=h/(1-2J'/3\Delta)$, and the system is stable only when
$3\Delta>2J'$, implying that our approximate mean-field treatment is 
only valid when $m_H>2J'/3$. It is also easy to show 
from \ (\ref{mfe}) and \ (\ref{spectrum}) that the 
Haldane gaps $m_H^{(m)} (m=0,\pm1)$ increase as a function of $h$, 
with increase in Haldane gap $\Delta{m}_H^{(0)}\sim2\Delta{m}_H^{(\pm1)}$, 
in qualitative agreement with experiment and numerical DMRG results.

   The impurity Hamiltonian $H_{im}$ generate an extra term in the
rotor Hamiltonian,
\begin{equation}
\label{himr}
H_{rot}^{(im)}={(h'-h)\over2}(n_0^z+L_0^z+n_1^z-L_1^z).
\end{equation}
As a result, the effective staggered magnetic field acting on the spin 
chain becomes position dependent, with $h(i)=h$ for $i\neq0,1$, and
$h(i)=(h+h')/2$ for $i=0,1$. Applying mean-field analysis as before, we 
obtain new mean-field equations for every site $i$ in the rotor model,
\begin{eqnarray}
\label{mfei}
J'(<n^z_{i+1}>+<n^z_{i-1}>) & = & h_{eff}(i)-h(i),   \\  \nonumber
<n^z_i> & = & \gamma(h_{eff}(i))({h_{eff}(i)\over3\Delta}),
\end{eqnarray}
which can be solved numerically. In addition, the $({h'-h\over2})L^z$ 
terms in \ (\ref{mfei}) gives rise to shift in the energies of the
$|m=\pm1>$ states in the local rotors at sites $i=0,1$ with
$E_{\pm1}-E_G=\Delta(1+\gamma(h_{eff}(i)))\mp{(h'-h)\over2\sqrt{3}}$.
   
\begin{figure}[ht]
\epsfxsize=4 in
\epsfbox{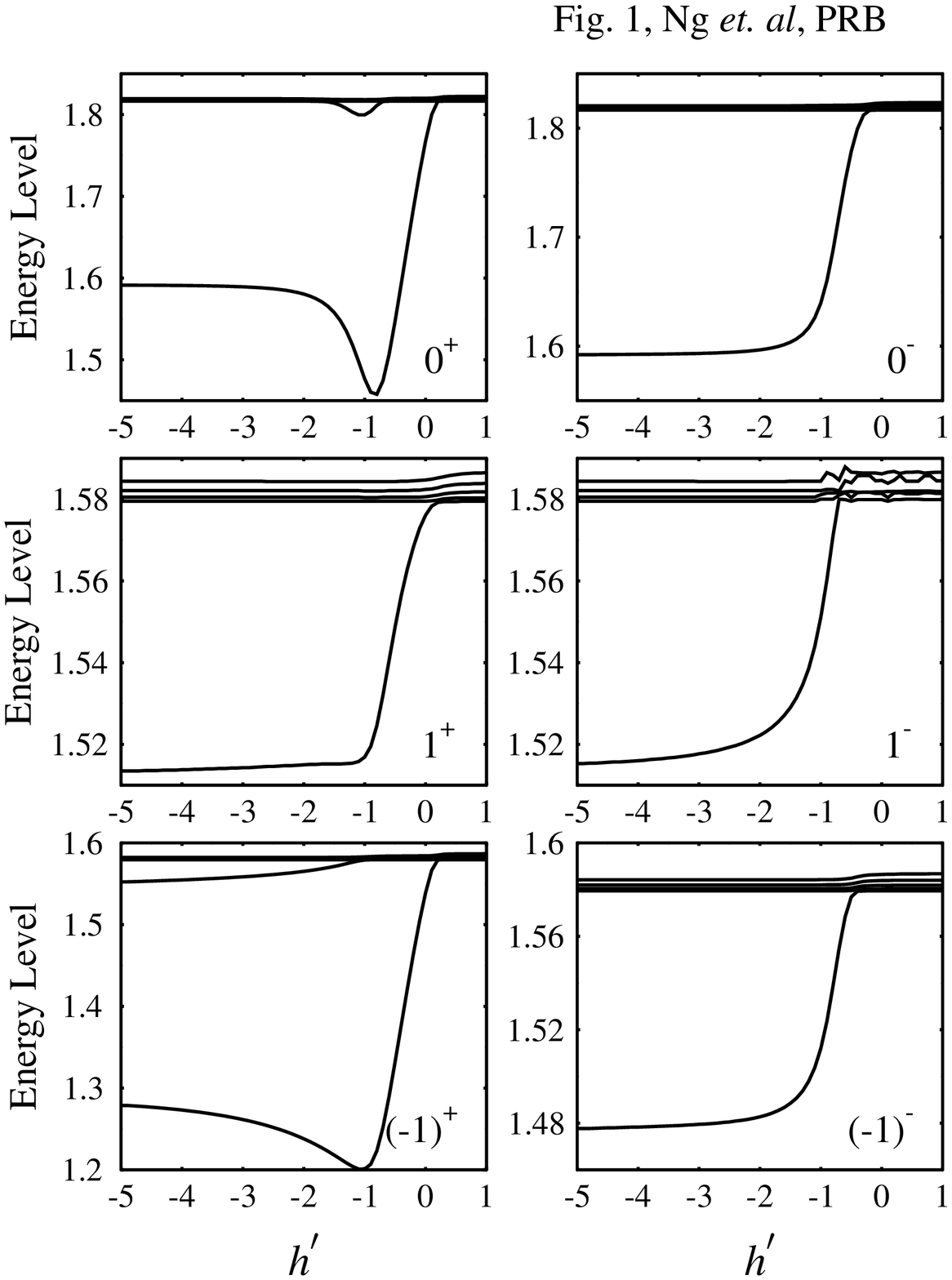}
\caption[]{
The low energy spectrum in the presence of single impurity
as a function of $h^{\prime}$ obtained 
in our mean-field theory with
$\Delta =1 $, $J^{\prime}$ = 1.47 and $h$ = 0.2.
}
\label{th1}
\end{figure}

     The effects of impurity magnetic field on the one magnon spectrum 
can be understood quite easily. The mean-field equations \ (\ref{mfei})
produce effective staggered magnetic field $h_{eff}(i)$'s smaller
than $h$ (assuming $h'\sim-h$) within a region of distance $\sim\xi$ 
around the impurity perturbed sites $i=0,1$, where $\xi$ is the spin-spin 
correlation length. As a result, the local magnitude of Haldane gap is
reduced (see Eq.\ (\ref{spectrum})) in both the longitudinal and
transverse channels and induces an attractive potential well of depth
$\sim\Delta{m}^{(m)}_H$ and range $\sim\xi$ for the one magnon states.
As a result bound states of magnon are formed around the impurity. The 
effective potential well formed around the impurity is stronger in the
longitudinal channel than the transverse channel as the effect of 
$h_{eff}$ on the Haldane gap is about two times larger in the 
longitudinal channel. Moreover, the energies of the $m=-1$ magnon 
states become lower than the $m=1$ states because of the additional
$-({h'-h\over2})L^z$ coupling in $H_{rot}^{(im)}$. In Fig. \ref{th1} we show 
the low energy spectrum obtained in our theory with $\Delta=1$, $J'=1.47$,
$h=0.2$ as a function of $h'$. Notice that in the presence of single 
impurity, the translational symmetry of the Hamiltonian is broken and the 
only symmetry left is the reflection symmetry upon the impurity site. As 
a result, the low-energy one-magnon spectrum of the system can be 
divided into six sectors labeled by $({S^z_{tot}})^{parity} = 0^{+}, 
0^{-}, 1^{+}, 1^{-}, (-1)^{+}, (-1)^{-}$, respectively where 
$S^z_{tot}(=m)$ is the $z$ component of total spin, and $+$ or $-$ 
denotes the parity of the wavefunctions. We see from the energy spectrum 
that spinwave bound state develops when $h'<0.2$, and there is roughly 
one bound state (except the $-1^+$ magnons at intermediate value of
$h'$) per sector. 

\begin{figure}[ht]
\epsfxsize=4 in
\epsfbox{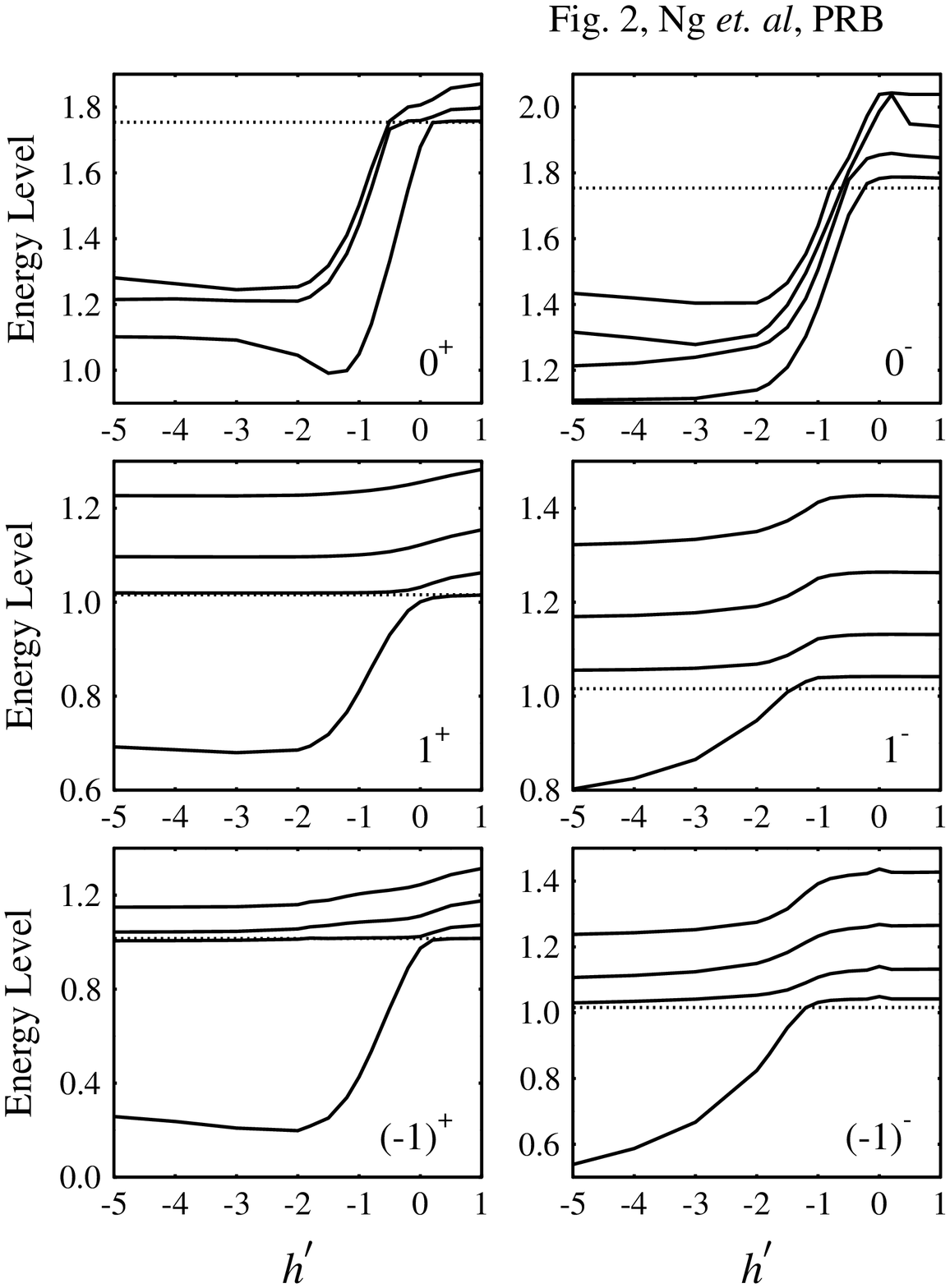}
\caption[]{
The the low energy excitation states 
of the six sectors for staggered magnetic field  $h$=0.2 
and chain length L=100,  the dotted line is the corresponding value
of the Haldane gap for $h$=0.2 without impurity.
}
\label{gap1}
\end{figure}

   To test our theory and to get more quantitative predictions of
the impurity effect we also carry out explicit DMRG calculations
for the low energy spectrum of Hamiltonian $H=H_{bulk}+H_{im}$. We
use the standard DMRG algorithm\cite{dmrg1,dmrg2} by targeting 
four states and keeping 400 optimal states, and have studied
chains with length up to $L=100$. We have fixed $h = 0.2(J)$ in our 
calculation, which seems to be an appropriate value for the $R_2BaNiO_5$
chains\cite{n2}. The value of $h^{\prime}$ is varied to study the 
effect of impurity on the low energy spectrum. The numerical results 
of the low energy magnon spectrum in the six $({S^z_{tot}})^{parity}$ 
sectors are shown in Fig. \ref{gap1}. Our results show that in the four sectors 
$1^{+}$, $1^{-}$, $(-1)^{-}$, $(-1)^{+}$ corresponding to the transverse 
modes, one bound state is induced in each sector when $h^{\prime}<0.2$, 
in agreement with our mean-field theory. The behavior of the bound 
state energy as a function of $h'$ is also qualitatively similar to 
those obtained in mean-field theory. In the $0^{+}$ and $0^{-}$ 
sectors which correspond to longitudinal modes, the situation is more
complicated. We find that more than one bound states emerge inside the 
Haldane gap in contrast to mean-field theory. Notice that although the 
number of bound states differs, the qualitative behavior of the lowest 
bound state energies as a function of $h'$ in the $0^+$ and 
$0^-$ sectors are qualitatively similar to mean-field results. 

  The reason why there exists only one bound state in mean-field theory
can be understood quite easily. Our mean-field theory is valid only
when $m_H>2J'/3$, corresponding to correlation length $\xi\sim1$ lattice
site whereas $\xi\sim6$ lattice sites in real spin one chains. The size 
of the effective attractive potential well is of order $\sim2$ lattice
sites in our theory, which is much smaller than the size of the
effective attractive potential well ($\sim12$ lattice sites) in real
spin one chains. It is therefore not surprising that the number of
bound states in real spin one chains is larger than one per sector but 
there exists only one bound state per sector in our mean-field theory.

\begin{figure}[ht]
\epsfxsize=\columnwidth
\epsfbox{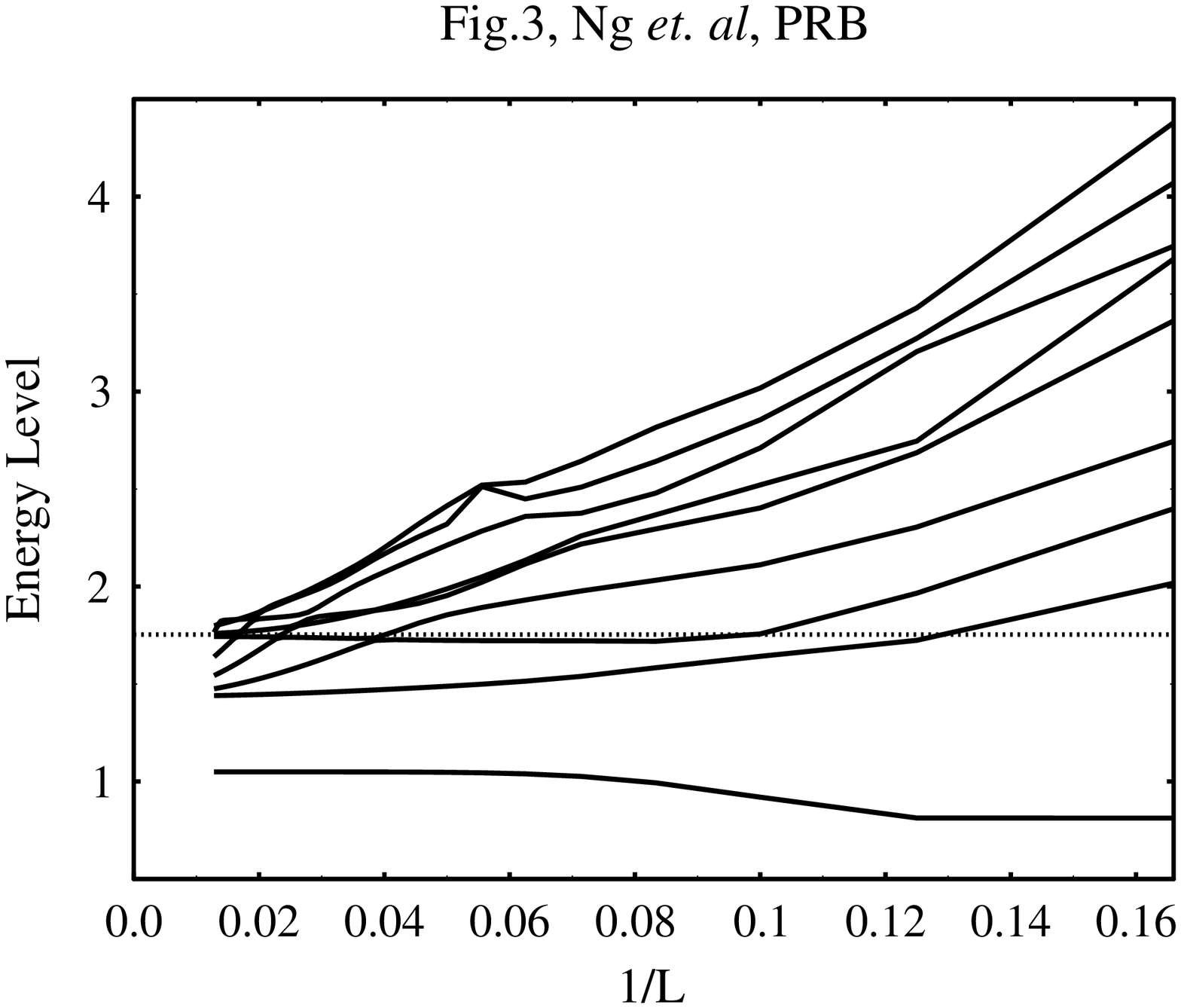}
\caption[]{
The chain length dependence of the lowest nine energy levels 
for $0^+$ sector for $h$=0.2 and $h^\prime$=-1.0, the dotted line
is the gap value at the thermodynamical limit.
}
\label{gap2}
\end{figure}

 To determine the number of the bound states in the longitudinal sectors, 
we carried out a more detailed calculation for $0^{+}$ sector by
targeting ten states and keeping 800 optimal states for $h^{\prime} = 
-1.0$ and $h = 0.2$, the chain length dependence of these states are 
shown in Fig. \ref{gap2}. We see from the figure that as the length 
of the chain increases, more in-gap bound states are formed. In fact, it 
is difficult to determine from our numerical result the exact number of 
bound states in this sector. 

\begin{figure}[ht]
\epsfxsize=4 in
\epsfbox{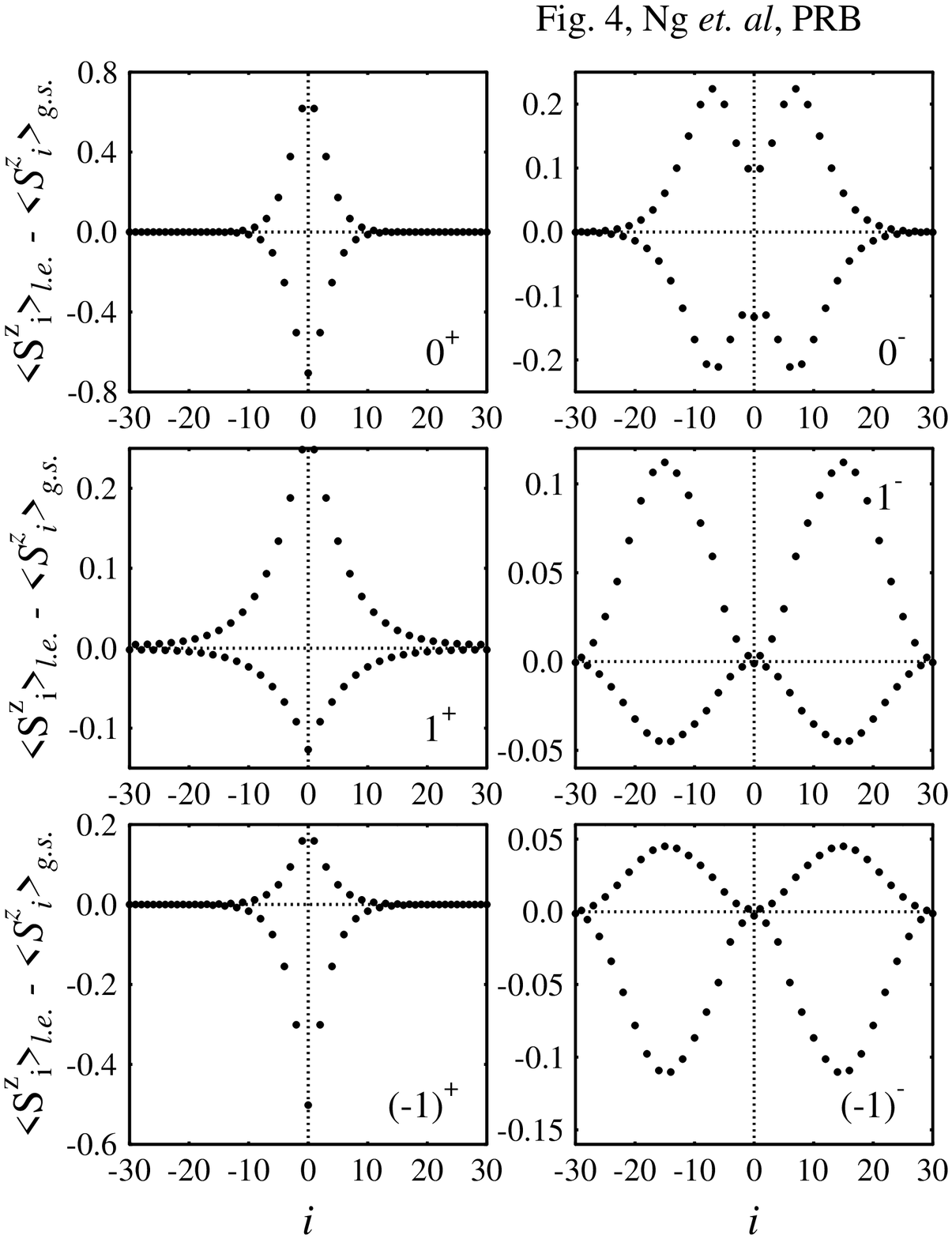}
\caption[]{
The magnetization density difference of the lowest
excitation state of each sector to the ground state,  
$<S^z_i>_{\it l. e.} - <S^z_i>_{\it g. s.}$  at the physicsl
regime $h'=-0.2$.
}
\label{mag1}
\end{figure}

In the physical regime $h'=-0.2$ which corresponds to non-magnetic 
impurity, we find that bound states exist in the $0^+, 0^-, 1^+$ and 
$(-1)^+$ sectors.  The bound state energies in the $0^+$ and $(-1)^+$
sectors are larger and are roughly $0.1\times{m}_H$, where $m_H$ are
the corresponding Haldane gap energies, whereas the bound state energies 
in the $0^-$ and $1^+$ sectors are much smaller ($\sim0.02\times{m}_H$. 
The existence of bound states at $h'=-0.2$ can be seen more clearly
from the magnetization density difference $<S^z_i>_{\it l. e. } - 
<S^z_i>_{\it g. s.}$ as a function of position $i$, which measures
the magnetization carried by the lowest excited state at each sector.  
The results are shown in Fig. \ref{mag1}. We can see clearly from the
figures that the first excited states of $0^+, 0^-, 1^+$ and 
$(-1)^+$ sectors are bound magnon states whereas the first excited states
of the $1^-$ and $(-1)^-$ sectors are bulk states. 

The bound states can be observed in neutron scattering experiments on 
high quality crystals as an effective reduction of Haldane gap by 
$\sim10$ percent at $T<T_N$ when non-magnetic impurities are introduced. 
They also show up as extra spectral weights with energy $\sim0.9m_H$ in
dynamic structure factor $S(q,\omega)$ at wave vectors $q$ away from $\pi$. 

 In summary, we have studied the effect of substituting $R$ (rare-earth 
ion) by non-magnetic ions in the spin-1 chain material $R_2BaNiO_5$. We
find in numerical DMRG calculation that magnon bound states appear at 
the impurity site. The bound states can be qualitatively understood by
a mean-field strong-coupling expansion treatment of the spin chain and
can be observed in neutron scattering experiments. 

This work is supported by the HKUGC under RGC grant HKUST6143/97P
and by Chinese Natural Science Foundation, J. Lou would like to thank
Dr. Shaojin Qin and Dr. Tao Xiang for useful discussion.

\end{document}